\documentclass[%
 reprint,
 superscriptaddress,
 amsmath,amssymb,
 aps,
 floatfix,
]{revtex4-2}

\usepackage[T1]{fontenc}
\usepackage[utf8]{inputenc}
\usepackage{ORCIDinREVTeX} % Add ORCID
\usepackage{graphicx}% Include figure files
\usepackage{dcolumn}% Align table columns on decimal point
\usepackage{bm}% bold math
\usepackage{booktabs}
\usepackage{tabularx}
\usepackage{hologo}
\usepackage{graphicx}
\usepackage[version=4]{mhchem}

\begin{document}

\preprint{arXiv:2301.13681}

\title{Potential of Monolayer Charge}

\author{Anna Stepanova}
%\email{anna.stepanova@lu.lv}
\orcid{0009-0008-1846-6809}
\affiliation{%
 Department of Chemistry, University of Latvia, Jelgavas iela 1, LV-1004 Riga, Latvia
}

\author{Heigo Ers}
%\email{heigo.ers@ut.ee}
\orcid{0000-0003-2005-5597}
\affiliation{%
 Institute of Chemistry, University of Tartu, Tartu 50411, Estonia
}

\author{Ritums Cepitis}
%\email{ritums.cepitis@ut.ee}
\orcid{0000-0002-6384-7589}
\affiliation{%
 Institute of Chemistry, University of Tartu, Tartu 50411, Estonia
}

\author{Michal Mal\v{c}ek}
%\email{michal.malcek@stuba.sk}
\orcid{0000-0002-1920-1123}
\affiliation{%
 Institute of Physical Chemistry and Chemical Physics, Faculty of Chemical and Food Technology, Slovak University of Technology in Bratislava, Radlinsk{\'e}ho 9, SK-812 37, Bratislava, Slovak Republic
}

\author{Vladislav B. Ivani\v{s}t\v{s}ev}%
%\homepage{http://www.doublelayer.eu}
\email{vladislav.ivanistsev@lu.lv}
\orcid{0000-0003-4517-0540}
\affiliation{%
 Department of Chemistry, University of Latvia, Jelgavas iela 1, LV-1004 Riga, Latvia
}%

\author{Iuliia V. Voroshylova}
\email{voroshylova.iuliia@fc.up.pt}
\orcid{0000-0003-2921-5155}
\affiliation{%
 REQUIMTE LAQV, Department of Chemistry and Biochemistry, Faculty of Sciences, University of Porto, 4169-007, Porto, Portugal
}

\date{\today}

\begin{abstract}
In this letter, we develop the concept of the potential of monolayer charge (PMC).
Its main purpose is to serve as the fundamental reference potential for studying charged interfaces. 
We estimate PMC values for interfaces between Au(111) surface and frisbee-shaped ions.
Density functional theory calculations suggest that increasing ion area shifts the PMC to an experimentally measurable potential range. 
To guide experimental verification, we have derived an analytical expression, which relates ion area, surface--ion distance, ionic charge, and the corresponding PMC value.
Further exploration of the PMC can enrich interfacial electrochemistry and reveal interfacial electrophysics as an independent field.
\end{abstract}

\maketitle

%%%%%%%%%%%%%%%%%%%%%%%%%
%%%%%%%%%%%%%%%%%%%%%%%%%

\paragraph*{Introduction}

\begin{figure}
\includegraphics[width=3.33in]{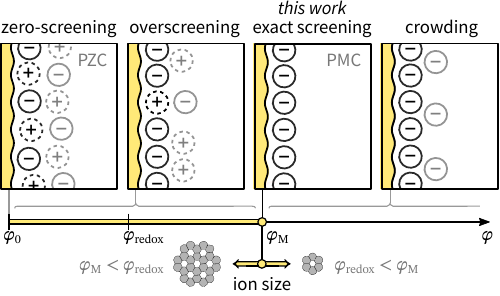}
\caption{\label{fig:cartoon} Schematic illustration of the interfacial structure: Potential of zero charge ($\varphi_{0}$); Overscreening potential range (where the surface charge is overcompensated by a monolayer of counterions, which is then balanced by the ions in the following layers); Potential of monolayer charge ($\varphi_{\mathrm{M}}$); and Crowding potential range (where the surface charge can not be compensated by a monolayer of counterions so that counterions also accumulate in the following layer).
$\varphi_{\mathrm{M}}$ stands as a milepost between overscreening and crowding.
By increasing the ion area, the $\varphi_{\mathrm{M}}$ value can be lowered below the redox potential ($\varphi_\mathrm{redox}$), \textit{i. e.} fitted into the electrochemical stability window.}
\end{figure}

Surface charge screening at interfaces is a central phenomenon in numerous research fields, from plasma physics through electrochemistry to colloid science \cite{block_double_1978,attard_recent_2001, wu_understanding_2022}.
The screening occurs within the electrical double layer (EDL) -- a nanometer-scale interfacial region that affects all interfacial processes and reactions. 
One of the main complications in studying screening is the EDL structural complexity -- a 3D distribution of charge carriers: ions and electrons \cite{kornyshev_three-dimensional_2014}. 
The complexity is most evident in the case of concentrated electrolytes such as ionic liquids (IL), where both short- and long-range interactions are crucial \cite{fedorov_ionic_2014}. 
Additionally, the finite size of ions, i.e. steric effects \cite{borukhov-steric-1997}, causes overscreening and crowding phenomena \cite{bazant_double_2011}, illustrated and defined in Fig.~\ref{fig:cartoon}.
Traditional methods to address this complexity, such as increasingly detailed and precise models, have yielded arguable results. 
For example, most advanced theories capture only the key features of overscreening and crowding \cite{bazant_double_2011,de_souza_interfacial_2020}. 
Meanwhile, computer simulations provide a more accurate description of these phenomena, yet they are hindered by computational cost and sampling quality \cite{jeanmairet_microscopic_2022}.

We suggest a different approach: Overcoming the EDL structural complexity by focusing on its simplest imaginable form at the \textit{potential of monolayer charge} (PMC) -- a~2D single layer of ions, which exactly screens the surface charge \cite{kirchner_electrical_2013}.
Molecular dynamics simulations suggest that PMC is a unique point between the overscreening and crowding mechanisms of surface charge screening \cite{ers_double_2022, voroshylova_ionic_2021}, as shown in Fig.~\ref{fig:cartoon}. 
Most recently, we concluded that large frisbee-shaped ions could be used to reach the PMC in electrochemical measurements \cite{karuIonicLiquidElectrode2024}.

In this Letter, we report the results of density functional theory (DFT) calculations demonstrating an inverse relationship between the PMC and the area of frisbee-shaped polycyclic heteroaromatic hydrocarbon ions. 
We propose an experimental setup to test the PMC concept in IL-based electrolytes and provide an analytical expression to guide the experimental verification.

\paragraph*{Concept}
---
PMC is a reference potential similar to the potential of zero charge (PZC; see Fig.~\ref{fig:cartoon}). 
The most remarkable feature of the EDL at the PMC is its dynamic and structural simplicity -- a virtually static monolayer of ions. 
This dense monolayer with a maximum packing density $\theta_{\mathrm{M}}$, exactly screens the surface charge density ($\sigma$): $\sigma_{\mathrm{M}} = - \theta_{\mathrm{M}} = -q/A$, where $q$ is the ionic charge, and $A$ is the surface area occupied by one ion. 
The ions and molecules that solvate the monolayer at one extra term ($\varphi_{\mathrm{solv}}$) to the surface--monolayer potential drop:
\begin{equation}
\label{eq:PMC}
\varphi_{\mathrm{M}} \approx -\frac{\theta_{\mathrm{M}}l}{\varepsilon_{0}}  + \varphi_{\mathrm{solv}},  
\end{equation}
\noindent where $l$ is the distance between the surface and ionic layer charge planes, and $\varepsilon_0$ is the vacuum permittivity.

As in the case of bare surfaces at the PZC, determination of $\varphi_{\mathrm{solv}}$ is challenging.
Still, PMC values can be defined using any potential scale ($U$) if referred to the PZC ($\varphi_0$ or $U_0$), as $\varphi_\mathrm{M} - \varphi_0 = U_\mathrm{M} - U_0$. 
The PZC and PMC are two reference potentials forming a self-consistent potential scale on which the ratio $\sigma/\sigma_\mathrm{M}$ equals 0 at the PZC and 1 at the PMC. The relation between $\sigma$ and $\varphi$ follows an asymptote \cite{nguyenIonicLiquidElectrode2025}:
\begin{equation}\label{eq:asymptot}
\frac{\sigma}{\sigma_{\mathrm{M}}} = \left( \frac{\varphi-\varphi_0}{\varphi_{\mathrm{M}}-\varphi_0} \right)^\alpha,
\end{equation}
\noindent where $\alpha$ is the potential-dependent scaling exponent \cite{voroshylova_ionic_2021}.
The $\varphi_{\mathrm{0}}$ matches the magnitude and sign of the $\varphi_{\mathrm{solv}}$ in Eq.~\ref{eq:PMC}. 

Below, we focus on the DFT-based estimation of PMC values for frisbee-shaped ions on the Au(111) surface, leaving other aspects for future studies.

\begin{figure}
\includegraphics[width=3.33in]{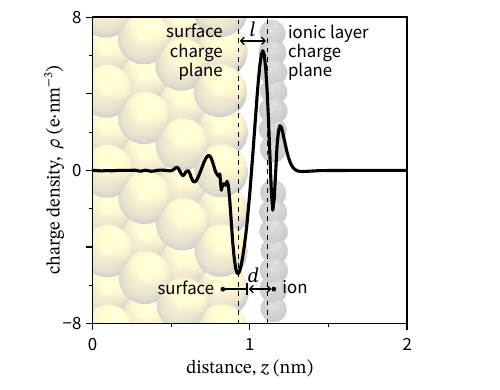}
\caption{Charge density \textit{vs.} distance dependence ($\rho(z)$; solid line), showing interfacial charge density fluctuations (perpendicular to the surface plane). Dashed vertical lines denote the electronic distance $l$ between the surface and ionic layer charge planes. Dots mark the positions of the surface and ion nuclei used to define the geometric distance $d$. The side view on the modeled interface is given in the background.}
\label{fig:ChgPlanes}
\end{figure}

\begin{figure}
\includegraphics[width=3.33in]{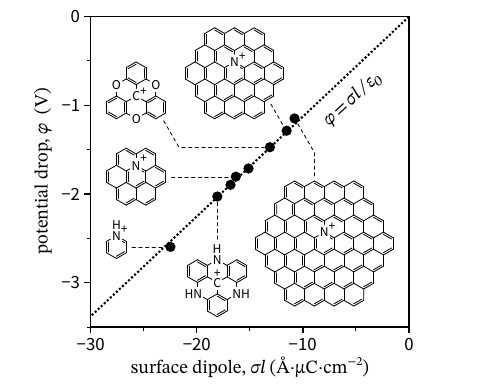}
\caption{\label{fig:PotDropvsSurfDip} Potential drop \textit{vs.} surface dipole dependence ($\varphi(\sigma l)$, data points), indicating a decrease of the absolute $\varphi$ values with an increase of ion area. Ions are illustrated with structural formulae.}
\end{figure}

\begin{figure}[!ht]
\includegraphics{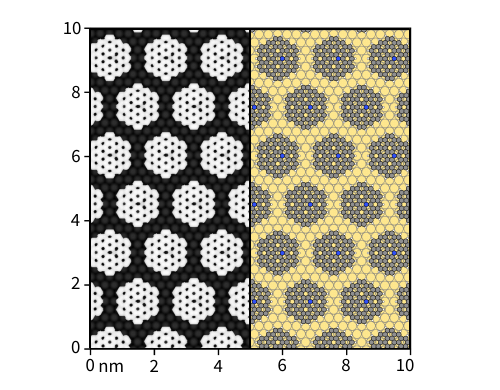}
\caption{\label{fig:STM} Left: Generated Tersoff--Hamann scanning tunneling microscopy image. Right: The top view on the modeled atomistic structure of the N-CCor$^{+}$ monolayer on the Au(111) surface.}
\end{figure}

\paragraph*{Methods}
---
Model interfaces consisted of a four-layer Au(111) slab (with the two bottom layers fixed in bulk positions) and commensurate ionic layer (packed as close as possible within a primitive hexagonal unit cell) of studied cations (Fig.~\ref{fig:PotDropvsSurfDip}): Pyridinium (Py$^{+}$), sesquixanthylium (TOTA$^{+}$), triazatriangulenium (TATA$^{+}$), azadioxatriangulenium (ADOTA$^{+}$), diazaoxatriangulenium (DAOTA$^{+}$),  azacoronenium (N-Cor$^{+}$), azacircumcoronenium (N-CCor$^{+}$), and azacircumcircumcoronenium (N-CCCor$^{+}$).
These models were relaxed to a maximum force of 0.05~eV$\cdot$\AA$^{-1}$.  
Then single-point calculations were conducted to evaluate quantities and properties, such as electron density (Fig.~\ref{fig:ChgPlanes}) and scanning tunneling microscopy image (Fig.~\ref{fig:STM}). 
Constrained DFT calculations were run to suppress surface--ion charge transfer \cite{melander_implementation_2016,meng_pragmatic_2022}. 
All calculations were carried out using the ASE, GPAW and NaRIBaS software \cite{enkovaara_electronic_2010,larsen_atomic_2017,roosnerutNaRIBaSScriptingFramework2018} with the PBE exchange-correlation functional \cite{perdew_generalized_1996} along with the D4 dispersion correction \cite{caldeweyher_extension_2020}, and the standard parameters \footnote{A 12 $\mathrm{\AA}$ vacuum layer above the ionic layer (at least 18 $\mathrm{\AA}$ between periodic images and dipole correction \cite{bengtsson_dipole_1999}), the grid-spacing of 0.182 $\mathrm{\AA}$, and a plane-wave basis with a 500~eV cut-off in spin-paired electron configuration were used.
The Brillouin zone was sampled using a Monkhorst--Pack grid \cite{monkhorst_special_1976}, where the number of k-points ($k$) in periodic directions was chosen so that the product $ka$ was greater than 70 k-point$\cdot\AA$, where $a$ is the length of the basis vector in a given direction.
Preliminary tests showed that the adsorption energy converges into $\pm$2.5~meV at the vacuum layer of 10 $\mathrm{\AA}$, the cut-off of 440~eV, $ka$ of 50 k-points$\cdot\mathrm{\AA}$.
The atomic regions were treated with the PAW formalism, and 11, 4, 5, 6, and 1 valence electrons were included for each Au, C, N, O, and H atom.
All optimizations (below 0.05~eV$\cdot$\AA$^{-1}$) were done with the quasi-Newton optimization algorithm \cite{byrd_limited_1995}. 
The lattice parameter of bulk gold was optimized (below 0.002~eV$\cdot\mathrm{\AA}^{-1}$) to 4.118~$\mathrm{\AA}$, using the exponential cell filter \cite{tadmor_mixed_1999}.
}.

\paragraph*{Analyses}
---
All presented data are openly accessible \footnote{Input scripts, proceeded results, and analysis scrips are available on Zenodo \cite{evaroosVilabtartuNaRIBaSPotential2023} which links to our GitHub repository: {https://github.com/vilab-tartu/NaRIBaS}{https://github.com/vilab-tartu/NaRIBaS}. The raw data can be regenerated using the input scripts or obtained from the authors upon request.}.
Surface charge density ($\sigma$), the distance between the charge planes ($l$), and the potential drop at the Au(111)--ion interface ($\varphi$) were calculated from the charge density: $\rho= - (\rho_{\mathrm{Au-ion}} - \rho_{\mathrm{Au}} - \rho_{\mathrm{ion}})$, where $\rho_{\mathrm{Au-ion}}$, $\rho_{\mathrm{Au}}$, and $\rho_{\mathrm{ion}}$ correspond to the all-electron densities of interface, Au(111) slab, and lone ion, respectively. 
The surface and ionic charge planes, illustrated in Fig.~\ref{fig:ChgPlanes}, were located as weighted averages using $\rho$ as weights. 
For this purpose, the $\rho(z)$ profile was divided into electrodic and ionic parts at $z_0$, where $\rho \approx 0$. The $\sigma$ values were evaluated as $\sigma = \int_{0}^{z_0} \rho(z) dz$.
The $\varphi$ values were calculated as the difference between the electrostatic potential at the boundaries of the simulation cell in the $z$ direction.
The electrostatic potential profile was calculated from $\rho(z)$ by Poisson's equation as in Ref. \cite{ers_grapheneionic_2020}.

\paragraph*{PMC and interfacial electrochemistry}
---
The PMC concept originates from molecular dynamics simulations of electrode--IL interfaces \cite{ers_double_2022, voroshylova_ionic_2021}, revealing that at the PMC, a single monolayer of ions can exactly screen the surface charge. 
Our recent work showed that the PMC values for 40 common IL ions fall outside the experimentally measurable electrochemical stability window of the corresponding ILs \cite{karuIonicLiquidElectrode2024}. 
In this work, we tested two sets of ions to validate the hypothesis that flat aromatic ions with a large area can be stable at PMC values. 
The first set contained four triangulenium ions of similar area \cite{bosson_cationic_2014}. 
Similar ions are well-known for their surface self-assembly and are considered candidates for being a molecular platform in molecular electronics \cite{wei_triazatriangulene_2014} and photonics \cite{jung_photoswitching_2012}.
The second set contained four azacoronenium ions of variable area.
The smallest ion, pyridinium, resembles the core of most common IL ions, such as alkylpyridinium and alkylimidazolium cations.

\begin{table}
\centering
\caption{Ionic charge and potential drop values from DFT ($q$, $\varphi$) and constrained DFT ($q_\mathrm{c}$, $\varphi_\mathrm{c}$) along with electronic ($l$) and geometric ($d$) distances between surface and ionic layers (see Fig.~\ref{fig:ChgPlanes}), and surface area occupied by one ion ($A$).}
\label{table:data}
\vspace{0.5cm}
\begin{tabularx}{\linewidth}{p{2.9cm}*{7}{>{\centering\arraybackslash}X}}
%\hline \hline\\[-8pt]
interface & $q$ & $\varphi$ & $l$ & $d$ & $A$ & $q_\mathrm{c}$ & $\varphi_\mathrm{c}$\\
          & [e] & [V] & [\AA] & [\AA] & [nm$^{2}$] & [e] & [V]\\

\hline
   Au(111)--N-CCCor$^{+}$ & 1.3   &  $-1.1$ &   1.9 & 2.1  &   3.6  & 1.0  &  $-0.8$ \\
   Au(111)--N-CCor$^{+}$  & 1.0   &  $-1.3$ &   1.9 & 2.1  &   2.6  & 1.0  &  $-1.3$ \\
   Au(111)--DAOTA$^{+}$   & 0.7   &  $-1.9$ &   1.9 & 2.0  &   1.2  & 1.0  &  $-3.0$ \\
   Au(111)--TATA$^{+}$    & 0.7   &  $-2.0$ &   1.9 & 2.0  &   1.2  & 1.0  &  $-3.0$ \\
   Au(111)--N-Cor$^{+}$   & 0.7   &  $-1.8$ &   1.8 & 2.1  &   1.2  & 1.0  &  $-3.1$ \\
   Au(111)--ADOTA$^{+}$   & 0.6   &  $-1.7$ &   1.9 & 2.0  &   1.2  & 1.0  &  $-3.1$ \\
   Au(111)--TOTA$^{+}$    & 0.5   &  $-1.5$ &   1.9 & 2.0  &   1.2  & 1.0  &  $-3.2$ \\
   Au(111)--Py$^{+}$      & 0.5   &  $-2.6$ &   2.0 & 1.8  &   0.7  & 1.0  &  $-5.9$ \\
%\hline \hline
\end{tabularx}
\end{table}

The established linear relationship in Fig.~\ref{fig:PotDropvsSurfDip} shows that as the area of the ions increases, the absolute potential drop decreases similarly to Eq. \ref{eq:PMC} as $\varphi = {\sigma l}/{\varepsilon_0}$, where $\sigma$ accounts for ion area and variable ionic charge due to the surface--ionic layer charge transfer.
Assuming the modeled packing density to be close to its maximum value, let us compare the calculated $\varphi$ values (Fig.~\ref{fig:PotDropvsSurfDip} and Table~\ref{table:data}) to the hypothetical $\varphi_\mathrm{M}$ for the studied Au(111)--ions interfaces.

For all interfaces, except for Au(111)--Py$^+$, $\varphi$ values are within the $-2$ to $-1$~V range, which is in the electrochemical stability window of ILs at electrodes such as Au(111) \cite{wallauer_differential_2013, mousavi_unbiased_2015}.
However, the ionic charges in Table \ref{table:data} indicate a significant charge transfer from the Au(111) surface to most ions.
Partial charge transfer of $\sim 0.2e$ is usual for ions in ILs \cite{lage-estebanez_self-interaction_2016}.
The reduction of $q$ from $+1e$ to much lower values indicates the electroreduction of the ions under the modeled conditions. 
Namely, the charge transfer over $0.3e$ for triangulenium family ions, Py$^+$ and N-Cor$^+$ also points to electroreduction.
That agrees with experimental data, showing that triangulenium ions exhibit one-electron reduction and remain adsorbed at the surface \cite{bosson_cationic_2014}.
Thus, we conclude that the studied triangulenium ions cannot form electrochemically stable monolayers at their theoretical PMC. 
Increasing the area and chemical stability of an ion \textit{via} functionalization, according to Eq.~\ref{eq:PMC}, is the way to reach realistically low absolute PMC values in an experiment.   
To our knowledge, triangulenium ions that are as flat as \mbox{N-Cor$^+$} yet have a much larger area have not been synthesized but could be acquired with evolving synthetic methods.

An intriguing case, worth of a separate study, is the anomalous (opposite) direction of charge transfer at the Au(111)--N-CCCor$^{+}$.
For N-CCor$^{+}$, the charge transfer is close to zero, which in terms of area and stability makes the Au(111)--N-CCor$^{+}$ interface an ideal candidate for experimental verification of the PMC concept.

In search of other candidates, let us compare the values in Table~\ref{table:data} for distances in Fig.~\ref{fig:ChgPlanes}: Between surface charge density planes based on the electron density ($l$); and between the nuclei positions of ions and surface atoms ($d$), shifted by one-half of an Au(111) interplanar spacing (see \cite{voroshylova_role_2019} and reference therein). 
The difference between $l$ and $d$ is small enough ($\sim 10\%$) to rely on geometric distances for the crude estimation of $\varphi_\mathrm{M}$ with Eq.~\ref{eq:PMC}.
For instance, for $d=2.1~\mathrm{nm}$ and $q=+0.8e$, PMC of $-1$~V corresponds to the ion area of 3~nm$^2$, which is comparable to the area occupied by one N-CCor$^{+}$ ion at the Au(111) surface in Fig.~\ref{fig:STM} (see also Table~\ref{table:data} for values).

Previous DFT studies of imidazolium and triangulenium ionic layers provide few details on interfacial charge distribution, but hint at the ionic nature of these ions at medium coverage of metal surfaces \cite{urushihara_theoretical_2015, jasper-tonnies_stability_2018}; 
For tetrafluoridoborate (BF${}_4^-$) anion layers, significant charge transfer and bond dissociation were reported \cite{ruzanov_thickness_2018}.
Differently from the named ions, azacircumcoronenium-like ions are explorable toy models with an ideal frisbee shape and high charge transfer resistance.

\paragraph*{PMC and interfacial electrophysics}
---
The enabling experimental studies of electrochemically stable interfaces near the PMC can lead to the establishment of a new field -- interfacial electrophysics.
Physical phenomena such as charge screening, electrowetting, electrostcriction, and electrofriction are commonly attributed to interfacial electrochemistry because they are chemically limited by charge transfer.
In this work, we have disabled the charge transfer by setting the ion charge ($q_\mathrm{c}$) equal to +1.0$e$ \textit{via} constrained DFT.
As shown in Table~\ref{table:data}, $\varphi_\mathrm{c}$ values for Au(111)--N-CCor$^{+}$ and Au(111)--N-CCCor$^{+}$ remained within the stability window of ILs, proving that similar ions are suitable candidates for exploring the above-listed physical phenomena at high absolute potentials near the PMC.
Curiously, similar azacoronenium ions are present in the interstellar space \cite{ricca_62_2021} but have not yet been synthesized as pure salts under laboratory conditions.
Encouragingly, reports on the synthesis of large aromatic ions have become more frequent. 
These ions maintain minimal $l$ in Eq.~\ref{eq:PMC} and reveal frisbee-to-oviform shapes \cite{w_laursen_synthesis_2001, wu_controllable_2009, wu_oxygen-_2009, yokoi_reversible_2018}.

While molecular dynamics simulations suggest that an ionic monolayer can exactly screen the surface charge at the PMC \cite{kirchner_electrical_2013}, it is unclear whether experimental observations will confirm this unique mechanism of screening. 
Overscreening in concentrated electrolytes and underscreening in diluted electrolytes have been observed in all known potential ranges, and it remains to be seen if nature circumvents exact screening at the PMC.
Even with the synthesis of stable, flat, and large ions and the atomic-level characterization of the EDL, it is possible that the monolayer alone cannot exactly screen the surface charge due to an undiscovered screening mechanism. 
Discovering a new screening mechanism that disproves the PMC concept would be as fruitful as confirming it.

One promising technique for studying monolayers at the PMC is \textit{in situ} scanning tunneling microscopy, offering direct imaging of the atomic scale structure and dynamics.\cite{wen_potential-dependent_2015,telychko_ultrahigh-yield_2021,ers_order_2022}  
The generated scanning tunneling microscopy image in Fig.~\ref{fig:STM} illustrates the kind of image that could be expected for \mbox{N-CCor$^{+}$} monolayer, exactly screening the surface charge around $-1.3$~V $\textit{vs.}$ PZC.
Supported by complementary research methods, this kind of imaging can provide valuable information about the structure and dynamics of the EDL and play an important role in establishing the PMC as the reference potential in interfacial electrochemistry and electrophysics.

\paragraph*{Conclusions}
---
This density functional theory-based study shows that at the potential of monolayer charge (PMC) large ions can form stable ionic monolayers, exactly screening the surface charge.
Our results suggest that the PMC values of such ions are experimentally measurable within the electrochemical stability window of common ionic liquids. 
We propose verifying the PMC concept with scanning tunneling microscopy and complementary techniques to establish the PMC as:
\begin{itemize}[leftmargin=*]
    \item the milepost potential between overscreening and crowding mechanisms of surface charge screening in concentrated electrolytes;
    \item the reference potential for guiding new directions in interfacial electrochemistry and electrophysics.
\end{itemize}

\paragraph*{References}

\bibliography{bibliography}

%\newpage

\begin{acknowledgments}
V.I., H.E., and R.C. were supported by the Estonian Ministry of Education and Research (TK210), Estonian Research Council grants PSG250, PSG249, and STP52; and the EU through the European Regional Development Fund (TK141, “Advanced materials and high-technology devices for energy recuperation systems”). V.I. received funding from the European Union’s Horizon 2020 research and innovation program under the Marie Sk\l{}odowska--Curie grant agreement no. 101031656. M. M. received financial support from the Slovak Grant Agency VEGA (contract No. 1/0324/24).
V.I. thanks Prof. L.M. Varela Cabo for discussing the concept presented.
H.E., V.I., and I.V. contributed equally to this work; A.S., R.C., and M.M. contributed to conceptualization and writing.
The authors declare no conflicts of interest.
\end{acknowledgments}

\end{document}